\documentclass[twocolumn]{aastex63}

\newcommand{\srcname}{Swift J0840.7$-$3516}
\newcommand{\swift}{{\it Swift}}
\newcommand{\nustar}{{\it NuSTAR}}
\newcommand{\nh}{N_\mathrm{H}}
\newcommand{\mbh}{M_\mathrm{BH}}

\shorttitle{Swift J0840.7$-$3516}
\shortauthors{Shidatsu et al.}

\begin{document}

\title{The Peculiar X-ray Transient Swift J0840.7$-$3516: an Unusual Low Mass X-ray Binary or a Tidal Disruption Event?}

\correspondingauthor{Megumi Shidatsu}
\email{shidatsu.megumi.wr@ehime-u.ac.jp}

\author[0000-0001-8195-6546]{Megumi Shidatsu}
\affil{Department of Physics, Ehime University, 
2-5, Bunkyocho, Matsuyama, Ehime 790-8577, Japan}

\author{Wataru Iwakiri}
\affil{Department of Physics, Faculty of Science and Engineering, Chuo University, 1-13-27 Kasuga, Bunkyo-ku, Tokyo 112-8551, Japan}

\author{Hitoshi Negoro}
\affil{Department of Physics, Nihon University, 1-8-14 Kanda-Surugadai, Chiyoda-ku, Tokyo 101-8308, Japan}

\author[0000-0002-6337-7943]{Tatehiro Mihara}
\affil{High Energy Astrophysics Laboratory, RIKEN, 2-1 Hirosawa, Wako, Saitama 351-0198, Japan}

\author[0000-0001-7821-6715]{Yoshihiro Ueda}
\affil{Department of Astronomy, Kyoto University, Kitashirakawa-Oiwake-cho, Sakyo-ku, Kyoto, Kyoto 606-8502, Japan}

\author[0000-0001-9656-0261]{Nobuyuki Kawai}
\affil{Department of Physics, Tokyo Institute of Technology, 2-12-1 Ookayama, Meguro-ku, Tokyo 152-8551, Japan}

\author[0000-0001-9307-046X]{Satoshi Nakahira}
\affil{High Energy Astrophysics Laboratory, RIKEN, 2-1, Hirosawa, Wako, Saitama 351-0198, Japan}

\author[0000-0002-6745-4790]{Jamie A. Kennea}
\affil{Department of Astronomy and Astrophysics, Pennsylvania State University, University Park, PA 16802, USA}

\author[0000-0002-8465-3353]{Phil A. Evans}
\affil{Department of Physics and Astronomy, University of Leicester, Leicester LEI 7RH, UK}

\author{Keith C. Gendreau}
\affil{Astrophysics Science Division, NASA Goddard Space Flight Center, Greenbelt, MD 20771, USA}

\author{Teruaki Enoto}
\affil{Extreme Natural Phenomena RIKEN Hakubi Research Team, RIKEN Cluster for Pioneering Research, 2-1 Hirosawa, Wako, Saitama 351-0198, Japan}

\author{Francesco Tombesi}
\affil{Department of Astronomy, University of Maryland, College Park, MD 20742, USA}
\affil{NASA/Goddard Space Flight Center, Code 662, Greenbelt, MD 20771, USA}
\affil{Department of Physics, Tor Vergata University of Rome, Via della Ricerca Scientifica 1, I-00133 Rome, Italy}
\affil{INAF Astronomical Observatory of Rome, Via Frascati 33, I-00078 Monte Porzio Catone, Italy}



\begin{abstract}
We report on the X-ray properties of the new transient \srcname, discovered with \swift/BAT in 2020 
February, using extensive data of \swift, MAXI, NICER, and \nustar. The source flux increased for 
$\sim 10^3$ s after the discovery, decayed rapidly over $\sim$ 5 orders of magnitude 
in 5 days, and then remained almost constant over 9 months. Large-amplitude short-term variations 
on time scales of 1--$10^4$ s were observed throughout the decay. 
In the initial flux rise, the source showed a hard 
power-law shaped spectrum with a photon index of $\sim 1.0$ extending up to 
$\sim 30$ keV, above which an exponential cutoff was present. The photon index 
increased in the following rapid decay and became $\sim 2$ at the end 
of the decay. A spectral absorption feature at 3--4 keV was detected in the decay. 
It is not straightforward to explain all the observed properties by 
any known class of X-ray sources.
We discuss the possible nature of the source, including a Galactic low mass X-ray binary with multiple extreme 
properties and 
a tidal disruption event by a supermassive black hole or a Galactic neutron star.

\end{abstract}



\section{Introduction} \label{sec:intro}

Observations of X-ray transients 
have provided valuable opportunities to study a huge variety of 
Galactic and extragalactic variable sources, from stellar 
objects including flaring stars, cataclysmic variables (CVs), 
X-ray binaries, and Gamma-ray bursts (GRBs), to supermassive 
black holes, including active galactic nuclei (AGN) and 
tidal disruption events (TDEs) of stars. 
In particular, the combination of all-sky monitoring 
and follow-up observations have deepened our knowledge of 
their behavior both in early and late phases after their 
emergence, and helped us to achieve critical information 
of their nature and physical mechanisms of energetic 
transient phenomena. 

\begin{figure*}[ht!]
\plotone{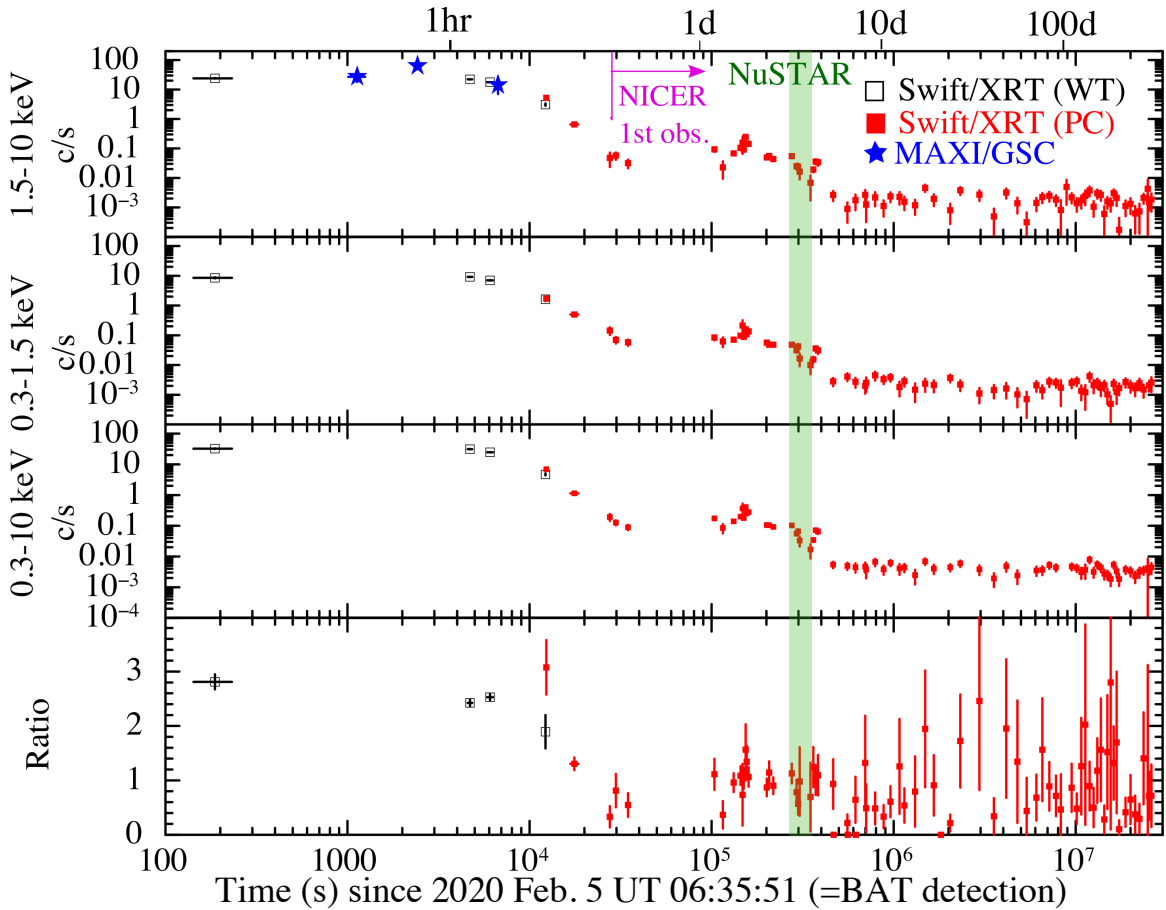}
\caption{\swift/XRT light curves in 1.5--10 keV,  
0.3--1.5 keV, and 0.3--10 keV, and the hardness ratio 
between 1.5--10 keV and 0.3--1.5 keV, from top to bottom. 
The black open and red filled squares represent the 
Windowed-Timing (WT) mode and the Photon-Counting (PC) 
mode data, respectively. They are binned in each snapshot 
before $4 \times 10^5$ s from the BAT detection ($T = 0$) 
and in each observation for the later time.
Errors in the abscissa and ordinate represent the start/end times 
of the bin from the mean photon arrival time, and 
the 1$\sigma$ errors of the count rates or the ratios, 
respectively. MAXI/GSC data are 
also plotted with blue stars (see \ref{sec:maxi} for the conversion 
to XRT count rates). \label{fig:longterm_lc_hid}}
\end{figure*}

\srcname~(also known as GRB 200205A) was discovered with 
{\it the Neil Gehrels Swift Observatory}~\citep[\it Swift;][]{geh04}
on 2020 February 5 UT 06:35 \citep{eva20a, ken20, osb20}, 
located at $\sim 4$ deg above the Galactic plane. 
Monitor of All-sky X-ray Image \citep[MAXI;][]{mat09} 
nova alert system \citep{neg16} also 
detected the source about 20 minutes after the \swift~detection 
\citep{niw20}. The MAXI/Gas Slit Camera (GSC) intensity 
increased from $\sim 70$ mCrab to $\sim 210$ mCrab 
in the 2--10 keV band from the first to second scan after 
the \swift~detection, and then rapidly decreased to $\sim 30$ 
mCrab in the third scan. 
Follow-up X-ray observations has been made with \swift~and the 
Neutron star Interior Composition ExploreR \citep[NICER;][]{gen16} 
59 and 16 times (as of 2020 December 10), respectively,
and once with the Nuclear Spectroscopic Telescope 
Array \citep[\nustar;][]{har13} in 2020 February 8--9. A 
possible periodicity of 8.96 s was reported 
by using the \swift/XRT data \citep{ken20}, 
but in following NICER observations 
no periodic variations were significantly 
detected \citep{iwa20}. 

Optical and UV counterparts 
were found in photometric observations \citep{lip20, mel20, mal20, maz20}. 
\citet{mal20} suggested that the source may be a Galactic source 
rather than a GRB because of the location and 
an unusually high optical/gamma-ray brightness ratio 
at early phases. No radio counterpart has been 
found so far; \citet{bor20} 
posed a 3$\sigma$ upper limit of 
18 $\mu$Jy at 7.25 GHz using Australia Telescope 
Compact Array (ATCA) on 2020 February 11. 

Despite the intensive follow-up observations, the nature 
of \srcname~is still unknown. In this article, we analyze 
X-ray data of the source obtained with \swift, MAXI, 
NICER, and \nustar~to understand its nature and radiation 
processes. We utilized HEAsoft version 6.26.1 for data 
reduction and analysis. Errors represent 90\% confidence 
ranges of single parameters unless otherwise specified.

\begin{figure}[ht!]
\plotone{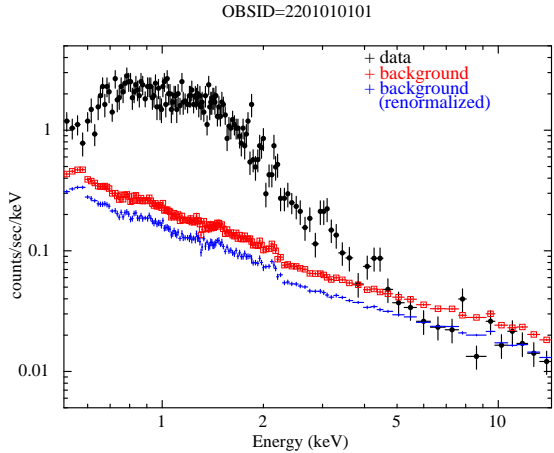}
\caption{(Black filled circles) time-averaged, folded spectrum 
produced from the first NICER observation (OBSID$=$2201010101). 
Background is not subtracted. (Red open squares) background spectrum 
generated with {\tt nibackgen3C50}. (Blue crosses) the same 
background spectrum renormalized so that its 10--15 keV count 
rate equals to that of the data.
\label{fig:compare_nibgd}}
\end{figure}

\begin{figure*}[ht!]
\plotone{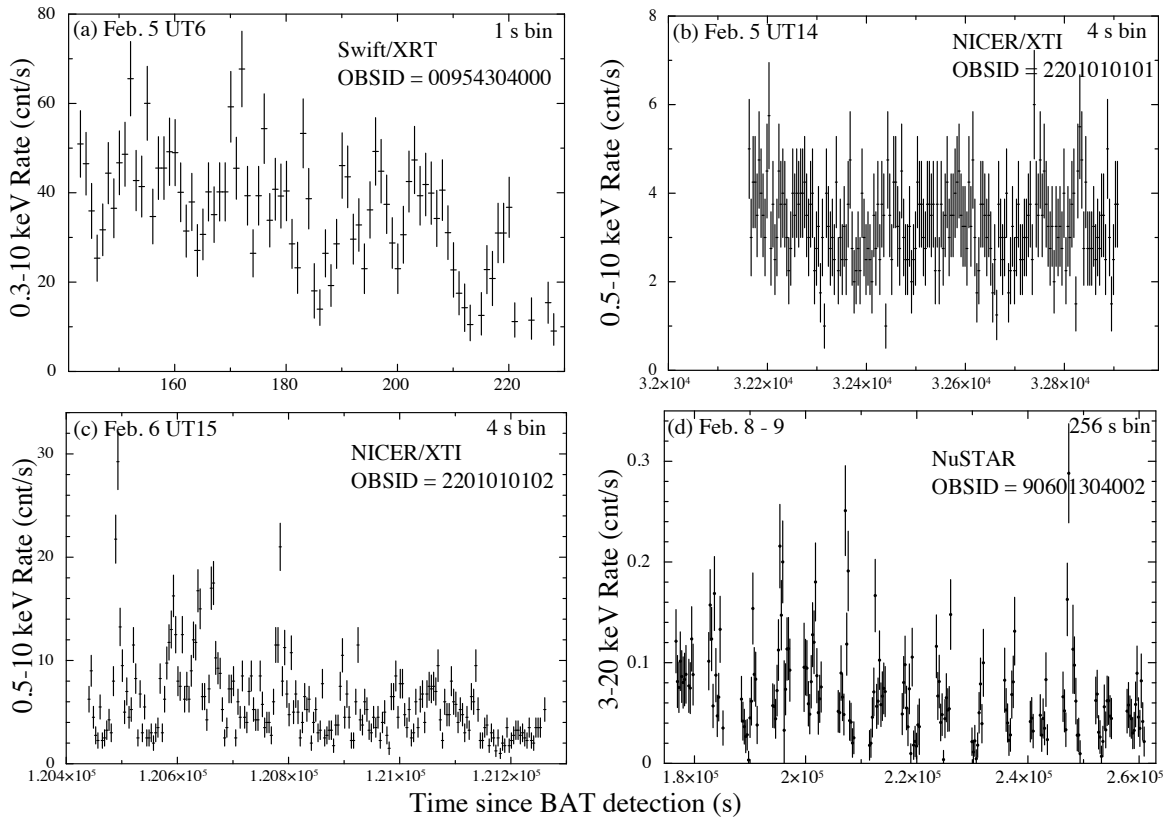}
\caption{Short-term light curves in different periods. Note 
that different energy ranges are adopted for \swift, NICER, and \nustar~(see the labels of the ordinate axes).
Background component is not subtracted in the NICER light curves. 
For clarity, the NICER light curves only cover the first 
continuous scans of the observations, while the other 
two include the entire periods of the observations. 
\label{fig:lc_short}}
\end{figure*}

\section{Observation and Data Reduction} \label{sec:data}

\subsection{\swift}
Since the first BAT detection at February 5 UT 06:35, 
\swift~observed the source almost every day 
in February and about once a week from the beginning of 
March to the end of November, 
with exposure times of 0.1--5 ks. We used all XRT data 
of the 59 observations.
Figure~\ref{fig:longterm_lc_hid} shows 
the \swift/XRT light curves and hardness ratio 
for the entire period, obtained via the web interface 
provided by the \swift~team\footnote{\url{https://www.swift.ac.uk/user_objects/}}
\citep{eva07, eva09}, in which we used 
the latest CALDB (version 2019 September 10) when 
this source was first observed with \swift.
The same interface was used to retrieve XRT spectra. 
We created time-averaged spectra for the individual 
observations until February 9. An exception is the second 
observation (OBSID$=$00954304001) performed in February 
5 UT 07-16, which was divided into each snapshot 
(i.e., each continuous exposure)
interval because of strong variations, although the last three 
intervals in UT 14-16 were merged due to low statistics. 
We combined the data for every $\sim$40 days from February 10 
to June 14, and merged all the data from June 15 and later,
to improve statistics of the spectra after the decay.

A time-averaged hard X-ray spectrum was also made from 
the BAT event-mode data taken at the first detection on 
February 5 (OBSID$=$00954304000). The BAT data were 
downloaded from HEASARC archive\footnote{\url{https://heasarc.gsfc.nasa.gov/cgi-bin/W3Browse/w3browse.pl}} and processed with the script {\tt batgrbproduct} 
included in HEAsoft, referring to the \swift/BAT 
Calibration Database (CALDB) released on 2017 October 16.
We adopted the default binning (binned in every 2 keV)  
to produce the spectrum.

\subsection{MAXI} \label{sec:maxi}
MAXI/GSC detected the source $\sim$20 minutes after the BAT 
detection and the subsequent two scans. We retrieved 
the GSC data in these individual scans via the 
on-demand system\footnote{\url{http://maxi.riken.jp/mxondem/}}. 
The GSC data are plotted in the top panel of
Fig.~\ref{fig:longterm_lc_hid}, together with the XRT data. 
They were converted to the XRT count rates in 1.5--10 keV 
through the HEASARC online tool WebPIMMS \citep{muk93}\footnote{\url{https://heasarc.gsfc.nasa.gov/cgi-bin/Tools/w3pimms/w3pimms.pl}}, 
using the best-fit parameters of the absorbed power-law model obtained 
from the time-averaged spectra in the corresponding 
scans (see Section~\ref{sec:spec_fit}). 
As noticed in the figure, MAXI uniquely  
observed the source around the flux peak of \srcname.

\subsection{NICER}

NICER started the first observation $\sim$8 hours after 
the BAT detection, and revisited the source every 1--few 
days until February 27, with exposure times of 0.4--10 ks. 
We used all the data taken before February 9, which 
have sufficient statistics for spectral and timing analysis. The NICER/XTI data were downloaded from 
HEASARC archive. They were reprocessed with the 
pipeline tool {\tt nicerl2} based on the NICER CALDB 
version 2020 February 2, before producing light curves 
and time-averaged spectra of the individual OBSIDs. 
The response matrix file {\tt nixtiref20170601v001.rmf} and ancillary 
response file {\tt nixtiaveonaxis20170601v003.arf} were adopted 
in spectral analysis. 
The background generator {\tt nibackgen3C50} version 4 was 
utilized to produce background spectra for each 
observation. The estimated 
background level was found to be overestimated in some OBSIDs, 
as shown in Figure~\ref{fig:compare_nibgd}. To absorb the 
deviation, we adjusted the normalizations of the 
background spectra of the individual OBSIDs so that the 
count rates in 10--15 keV 
(where background counts are dominant) are equal 
between the actual data and the generated background spectra.  
The renormalization factors are within the range 
of 0.9--1.4. We have found that this treatment 
only slightly changes the results of the spectral fits 
in the following sections (by $< 15$\% for the flux
and $<5$\% for the other parameters) 
from those obtained by using the original background 
spectra and that this does not affect the conclusions of the paper.

\subsection{\nustar}
The \nustar~observation was performed from February 
8 UT 07:18 to 9 UT 06:46 with an exposure time 
in the normal observing mode of 42 ksec (after standard 
screening to produce cleaned event data was applied).
The \nustar~data were retrieved from the HEASARC archive and 
reprocessed through {\tt nupipeline} and the latest 
\nustar~CALDB as of 2020 February 5. \nustar~light curves,   
the time-averaged spectrum, and its response files were 
created with {\tt nuproducts}. The source and background 
region were defined as circular regions with a radius 
of $50"$ centered on the target position, and with a 
$80"$ radius in a blank sky area on the same chip, 
respectively. 

\section{Analysis and Results}
\subsection{Long-term Light Curve}
As shown in Fig.~\ref{fig:longterm_lc_hid}, the source  
initially increased its intensity and reached the peak at 
$T \sim 10^3$ s (where $T$ is the time since the BAT trigger). 
It then exhibited a very rapid decay, decreasing its 
1.5--10 keV intensity by $\sim$ 5 orders of magnitude in 
$< 10^6$ s. 
In the decay, the source showed a plateau phase 
from $T \sim 3 \times 10^4$ s to $T \sim 1 \times 10^5$ s. 
In the early decline phase before the plateau, the 
hardness ratio, calculated by dividing the 1.5--10 
and 0.3--1.5 keV count rates, rapidly decreased, 
indicating spectral softening. After the end of the decay 
at $T \sim 5 \times 10^5$ s, 
the source showed flat light curves in all energy ranges, 
with a non-periodic variation by a factor of $\sim$ 2 (see 
the third panel of Fig.~\ref{fig:longterm_lc_hid}).

\begin{figure}[ht!]
	\plotone{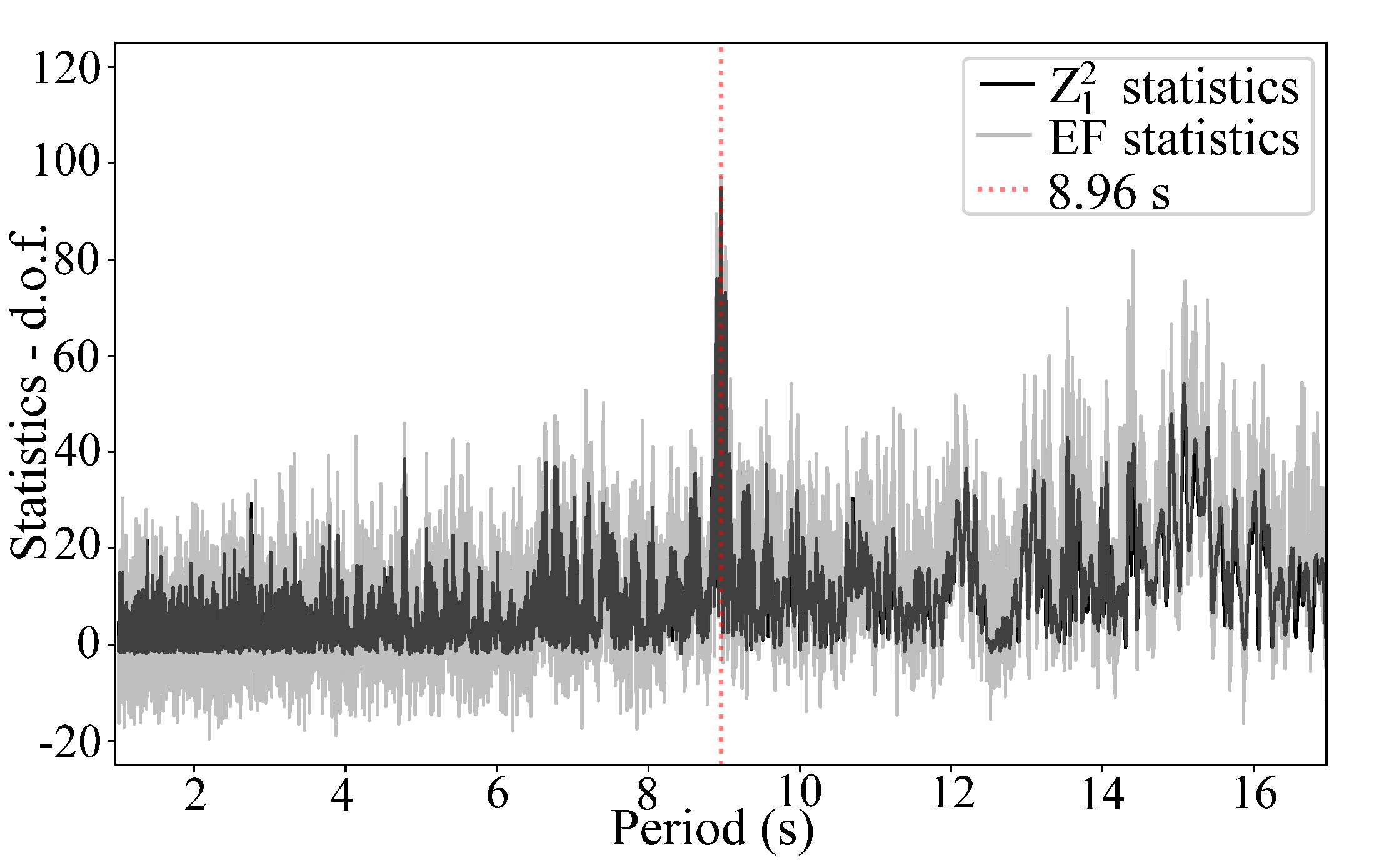}
\caption{Periodogram of \swift/XRT WT-mode data 
taken on 2020 February 5, with an exposure time of 765 s. 
Analysis using $Z^2_1$ search \citep{buc83} 
and e-folding techniques are over-plotted (black and gray, respectively). 
A clear peak is seen at $P=8.96$~s (red dotted line). This represents 
the only dataset in which this period is detected.
\label{fig:powspec_sw}}
\end{figure}

\begin{figure}[ht!]
\plotone{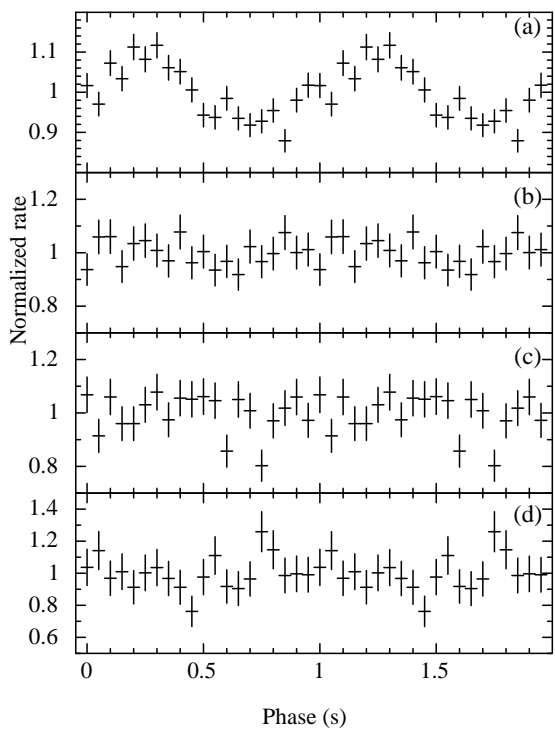}
\caption{Light curves in Fig.~\ref{fig:lc_short}(a)--(d) folded by 
8.96 s, normalized by their averaged count rates. The \swift/XRT 
data of OBSID$=$00954304001 were added to (a) and the whole observation periods were used in (b)--(d). 
\label{fig:folded_lc}}
\end{figure}

\begin{figure*}[ht!]
\plottwo{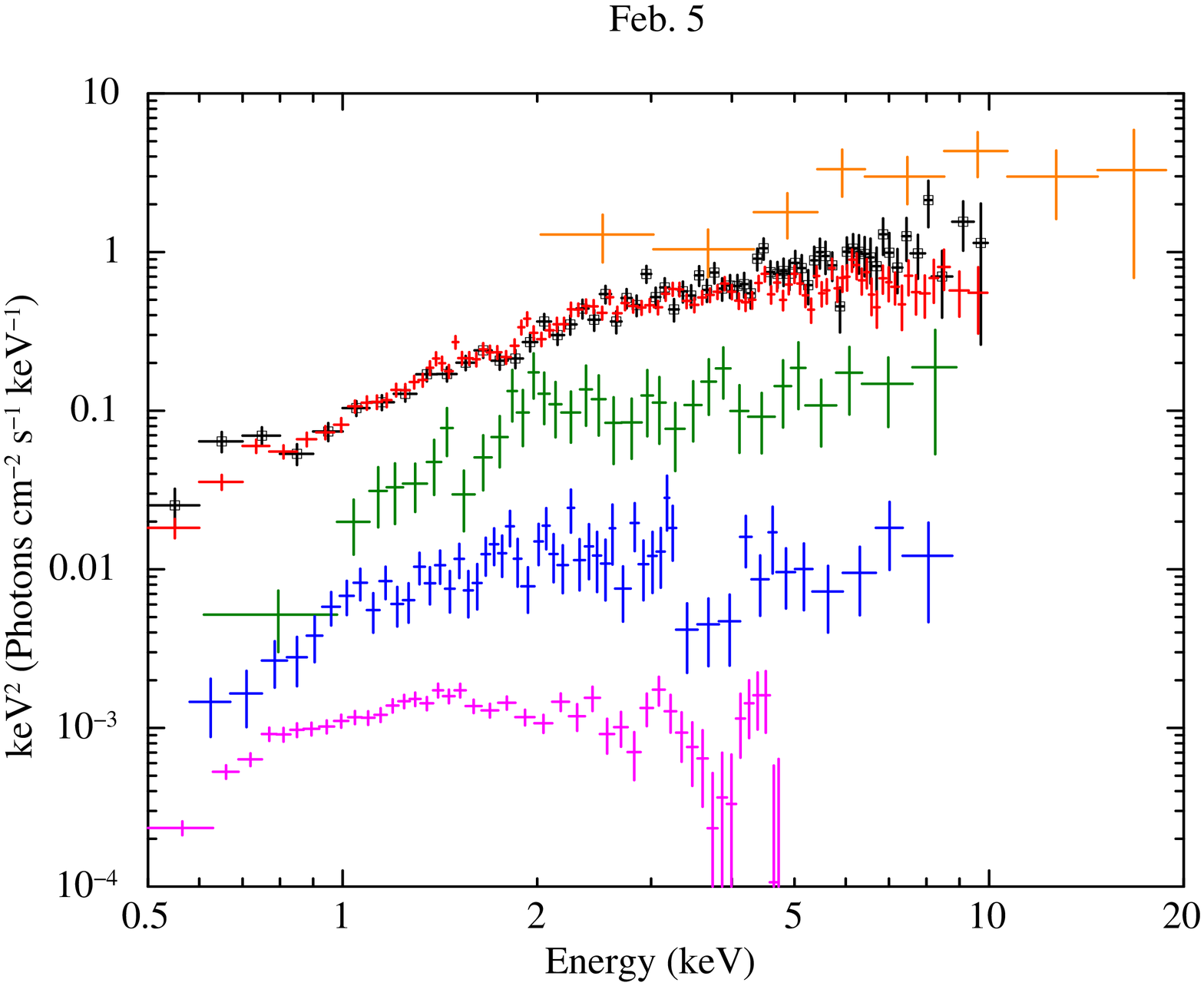}{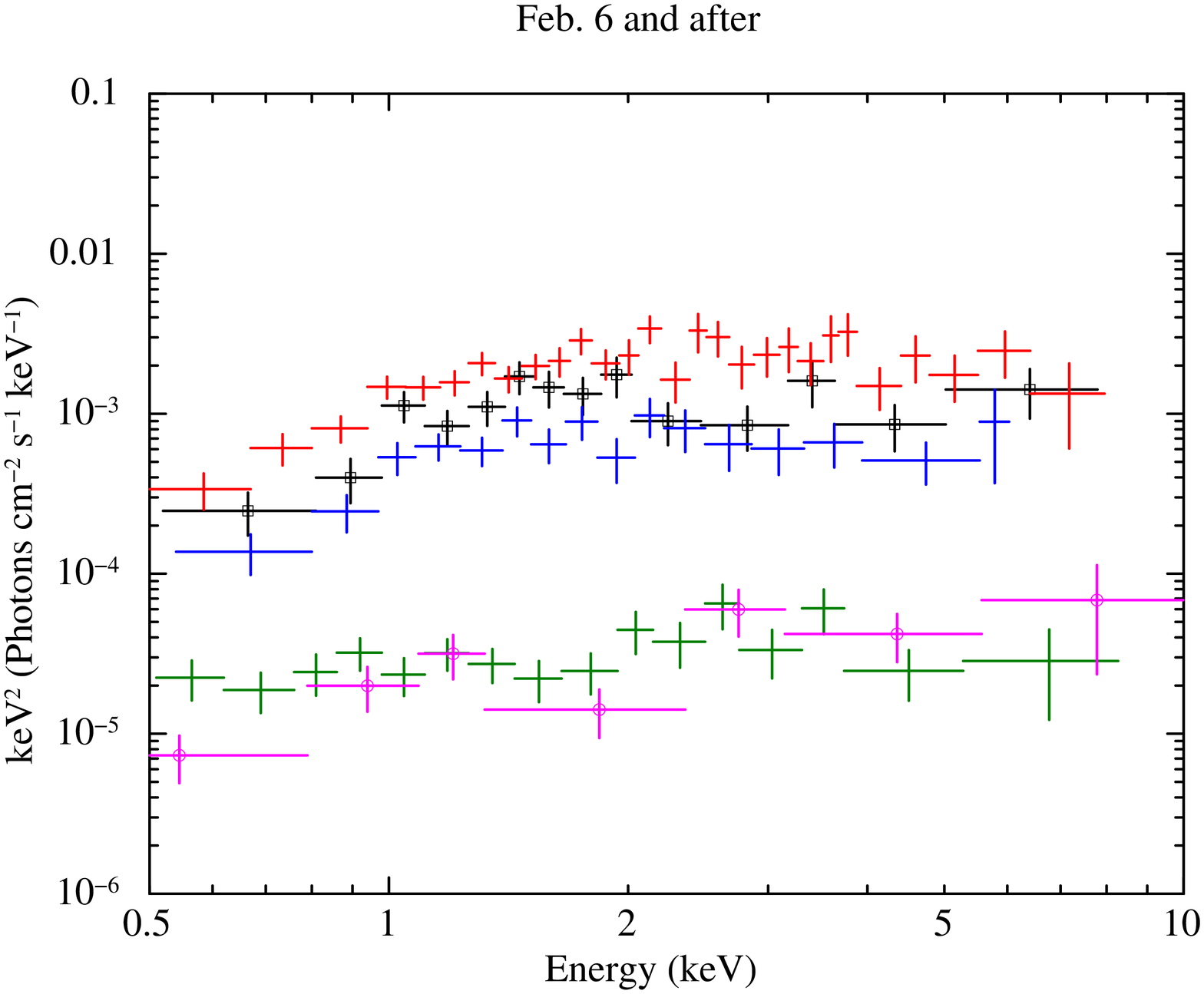}
\caption{Representative spectra on 2020 February 5 (left) and after (right), 
unfolded by using a single power-law model with a photon index of 2. 
(Left) The data of the first \swift/XRT observation in $T=$ (1.4--2.3) 
$\times 10^2$ s 
are presented with black open squares. The others are the MAXI/GSC spectrum 
obtained in $T=$ (2.3--2.5) $\times 10^3$ s (orange), \swift/XRT spectra 
in (4.5--4.8) $\times 10^3$ s (red), 
(1.22--1.24) $\times 10^4$ s (green), 
and (1.8--1.9) $\times 10^4$ s (blue), 
and NICER/XTI spectrum in (2--5) $\times 10^4$ s 
(magenta), from top to bottom. 
(Right) \swift/XRT spectra obtained 
in $T=$ (1.0--1.3) $\times 10^5$ s (black open squares), 
(2.3--2.5) $\times 10^5$ s (red), 
(2.7--3.0) $\times 10^5$ s (blue), 
and (4--37) $\times 10^5$ s (green), and 
(8--11) $\times 10^6$ s (magenta open circles). Note that 
some of the spectra were obtained by averaging multiple 
intermittent observations (see Fig.~\ref{fig:longterm_lc_hid}).}\label{fig:spec}
\end{figure*}

\subsection{Short-term Variability}
Figure~\ref{fig:lc_short} displays examples of 
short-term light curves at different periods. 
Strong variation 
on time scales of 1--$10^3$ s can be seen 
throughout the decay. The X-ray intensity 
varied by a factor of several to 
even more than one order of magnitude 
(see Fig.~\ref{fig:lc_short}d), although the 
amplitude became somewhat weaker around the end 
of the initial decline, $T \sim 3 \times 
10^4$ s (Fig.~\ref{fig:lc_short}c). We note that 
the strong variability was also clearly seen 
in the MAXI scan around the flux peak 
($T \sim 10^3$ s). 

We searched for periodicity using the available 
\swift/XRT, NICER, and \nustar~data, after applying 
barycentric correction for the target position 
(RA, Dec)(J2000) $=$ ($08^\mathrm{h}40^\mathrm{m}40^\mathrm{s}.84$, 
$-35^\circ16'24''.8$) with the ftool {\tt barycorr}, based on the 
JPL planetary ephemeris DE-200. Adopting the $Z^2_1$ \citep{buc83} 
and epoch-folding techniques, we found a peak at 8.96s in the 
periodogram of the first \swift/XRT WT-mode observations 
(OBSID$=$00954304000 and 00954304001; Fig.~\ref{fig:powspec_sw}),
which is consistent with the report by \citet{ken20}. 
The chance probability of this 8.96s period was 
$< 10^{-10}$ based on the $Z^2_1$ method. 
No significant periodicity was 
detected in the other data. In Figure~\ref{fig:folded_lc} 
we plot the light curve in Fig.~\ref{fig:lc_short} 
folded by 8.96 s using the ftool {\tt efold}. 

Using the light curves, we also searched for possible 
Type-I X-ray bursts, which are characterized by a short rise time of 
$\lesssim 10$ s and a subsequent longer decay for a few ten 
seconds or more \cite[e.g.,][]{lew93}. However, no evidence 
of the X-ray bursts were found in any of the \swift, NICER, 
and \nustar~data. Specifically, the flare seen in the 
\nustar~light curve (Fig.~\ref{fig:lc_short}d)
at $\sim 2.48 \times 10^5$ s from the BAT detection was found to have 
a $\sim 200$-s rise time, by using a light curve with shorter time bins, 
and hence is unlikely to be a Type-I X-ray burst.

\subsection{Analysis of X-ray Spectrum} 

\subsubsection{Overall Spectral Evolution}
\label{sec:spec_fit}

Figure~\ref{fig:spec} presents representative soft 
X-ray spectra obtained with the \swift/XRT, MAXI/GSC, 
and NICER/XTI at different epochs. The spectrum was 
hardest at the initial, brightest phase 
($T \lesssim 10^4$ s), then rapidly softened, 
and finally became almost flat in the 
$E F_\mathrm{E}$ form. A hint of an absorption 
feature is seen at 3--4 keV in the \swift~and NICER spectra 
taken in $T = 2 \times 10^4$--$5 \times 10^4$ s 
(blue and magenta spectra in the  Fig.~\ref{fig:spec} left panel), 
which we look into in Section~\ref{sec:spec_abs}.

\begin{figure}[ht!]
\plotone{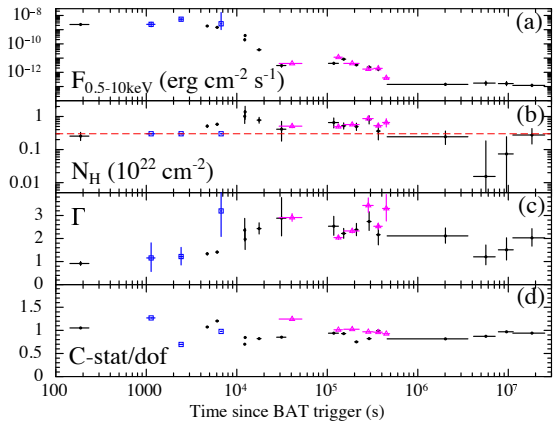}
\caption{Time variations in the best-fit parameters 
of the absorbed power-law model: (a) unabsorbed 0.5--10 keV 
flux, (b) $\nh$, (c) $\Gamma$, (d) C-statistic/(degree of freedom). 
\swift, MAXI, and NICER data are presented in black circles, blue squares,  
magenta triangles, respectively. The red dashed line in the panel 
(b) shows the total Galactic column ($\nh = 3 \times 10^{21}$ cm$^{-2}$).  
\label{fig:spec_par}}
\end{figure}

To quantify the overall spectral behavior, we 
applied an absorbed power-law model to each spectrum.
Throughout the spectral analysis, we used the 0.5--10 keV, 
2--20 keV, and 1.0--10 keV ranges of the \swift, MAXI 
and NICER data, respectively. C-statistic \citep{cas79} 
was employed because many of the spectra 
have relatively low statistics. 
The {\tt TBabs} model was adopted 
for the absorption component, with the solar 
abundance table given by \citet{wil00}. 
The equivalent hydrogen column density $N_\mathrm{H}$ 
was allowed to vary in \swift~and NICER data, while 
fixed in the MAXI data at $3 \times 10^{21}$ cm$^{-2}$, 
which is the Galactic column in the 
direction of \srcname~estimated with the tool 
{\tt nh} in HEAsoft. This is because MAXI 
is not sensitive to the soft X-rays below 2 keV
and difficult to constrain $\nh$.
We note that  
the free parameters obtained from the MAXI spectra remained  
consistent within their 90 \% confidence ranges when 
$\nh = 0$ and $1.0 \times 10^{22}$ cm$^{-2}$ were adopted. 
When $\nh$ is allowed to vary, it is not constrained at all 
and only the lower limits of $\Gamma$ and unabsorbed fluxes 
were obtained. 

\begin{deluxetable}{ccCC}[ht!]
\tablecaption{Best-fit 
parameters of absorption components. \label{tab:pars_abs}}
\tablecolumns{4}
\tablenum{1}
\tablewidth{0pt}
\tablehead{
\colhead{Parameter} & \colhead{Unit} &
\colhead{\swift} & \colhead{NICER} 
}
\startdata
\multicolumn{4}{l}{(1) {\tt TBabs*edge*powerlaw}} \\ 
$E_\mathrm{edge}$ & keV & 3.3 \pm 0.1 & 3.4^{+0.6}_{-0.2} \\ 
$\tau$ & & 1.43 \pm 0.7 & 0.52^{+0.78}_{-0.51} \\ 
C-stat/d.o.f & & 204/263 & 316/326 \\ \tableline 
\multicolumn{4}{l}{{(2) \tt TBabs*(gauss+powerlaw)}} \\ 
$E_\mathrm{cen}$ & keV &  3.7^{+0.2}_{-0.1} &  3.9^{+0.1}_{-0.2}   \\
$\sigma$ & keV & 0.2^{+0.2}_{-0.1} & < 0.36 \\ 
$N_\mathrm{gau}$\tablenotemark{a} & & 5^{+2}_{-3} \times 10^{-4} & 1.4^{+1.7}_{-1.1} \times 10^{-5} \\
EW & keV & 0.5^{+0.3}_{-0.2} & 0.24^{+0.29}_{-0.18} \\
C-stat/d.o.f & & 206/262 & 314/325 \\ \tableline 
\multicolumn{4}{l}{{\tt TBabs*powerlaw} (continuum)} \\ 
C-stat/d.o.f & & 276/265 & 319/328 \\ \tableline 
\enddata
\tablenotetext{a}{Absolute value of the normalization of {\tt gauss}, defined as the total photon flux in the absorption line, in units of photons cm$^{-2}$ s$^{-1}$.}
\end{deluxetable}

\begin{figure*}[ht!]
\plottwo{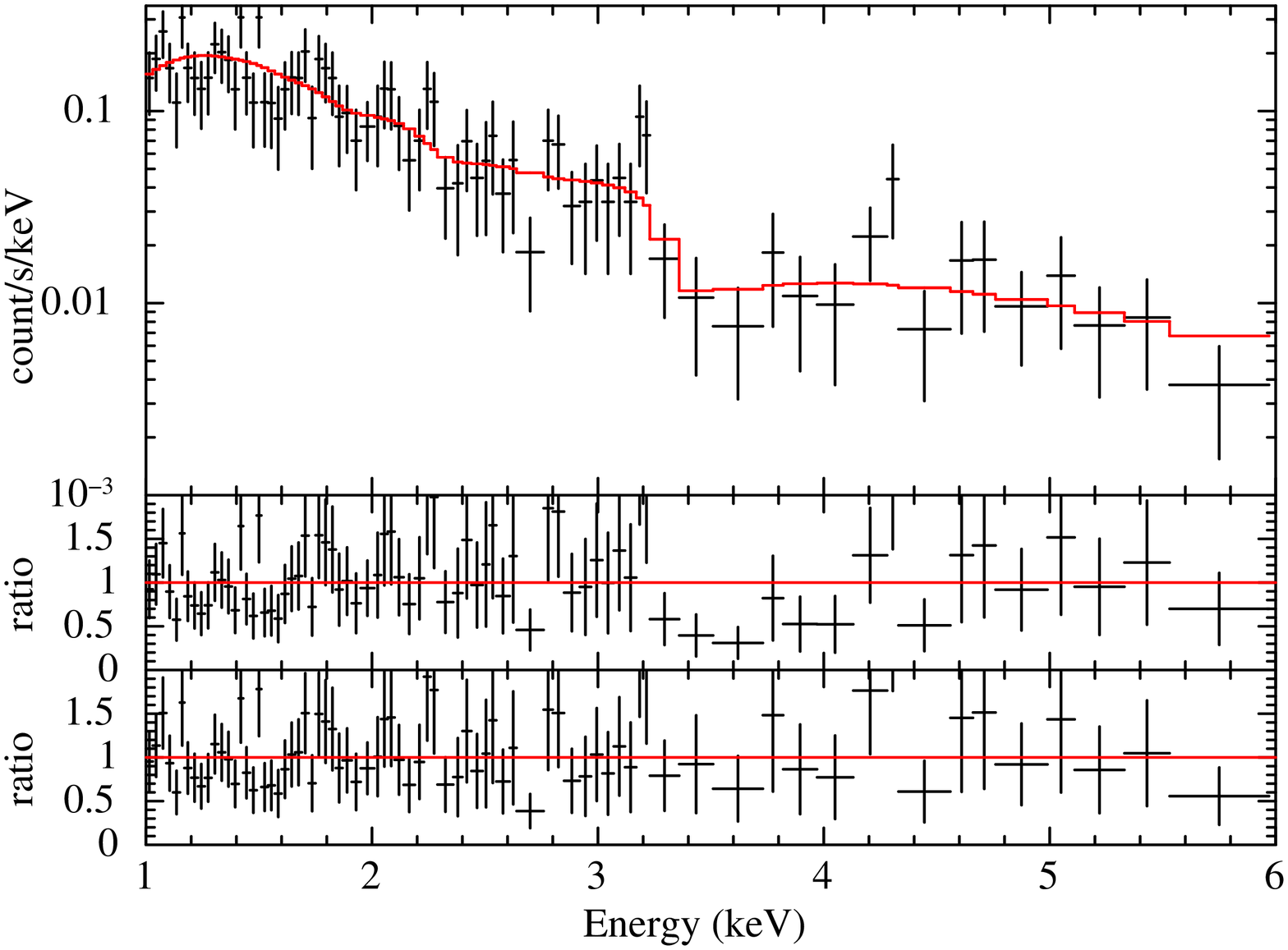}{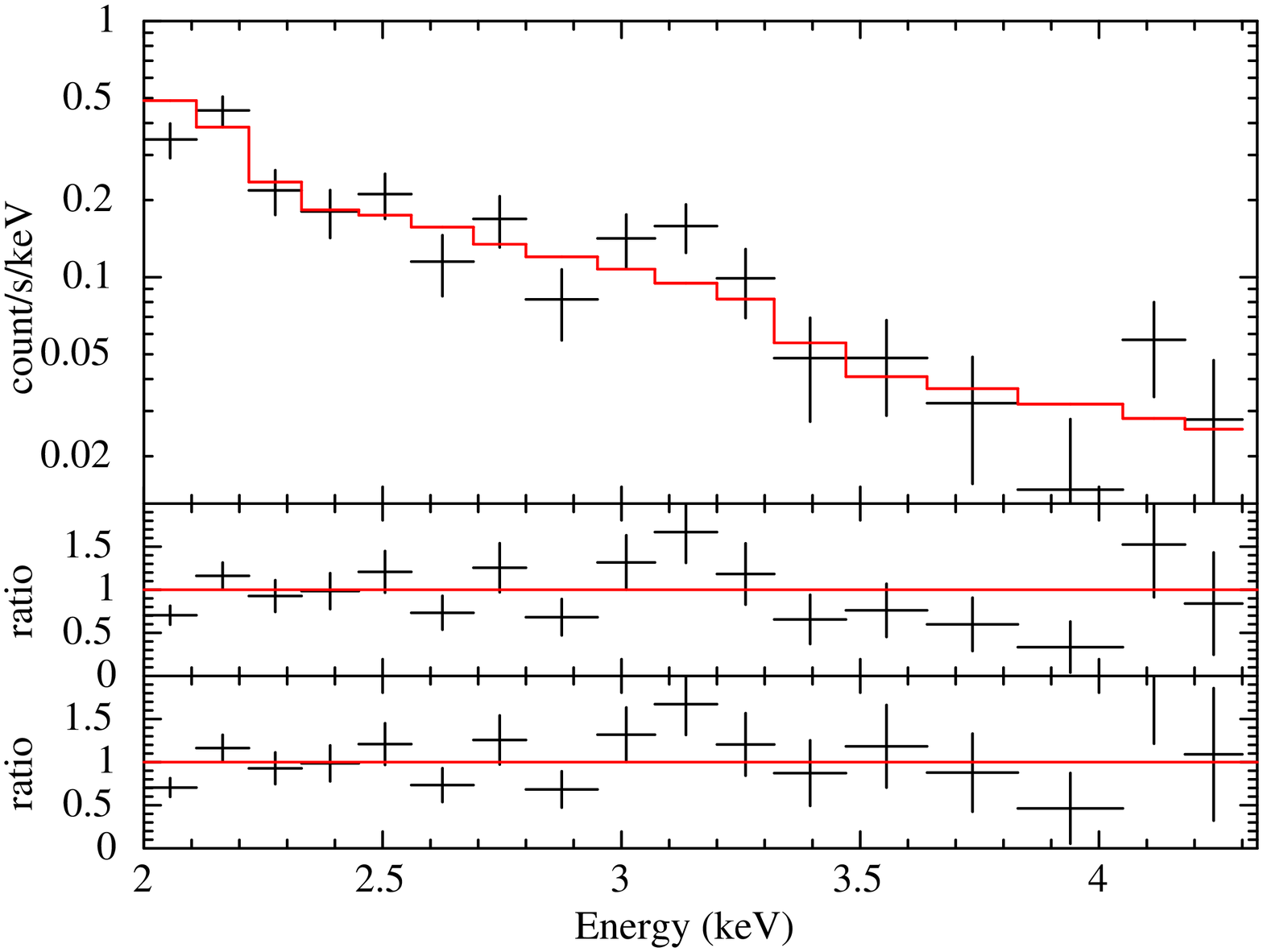}
\caption{\swift/XRT (left) and NICER (right) spectra and the best-fit 
{\tt TBabs*edge*powerlaw} model. The data versus model ratios for the 
models excluding and including {\tt edge} are plotted in the middle 
and bottom panels, respectively. \label{fig:spec_abs}}
\end{figure*}

\begin{figure*}[ht!]
\plotone{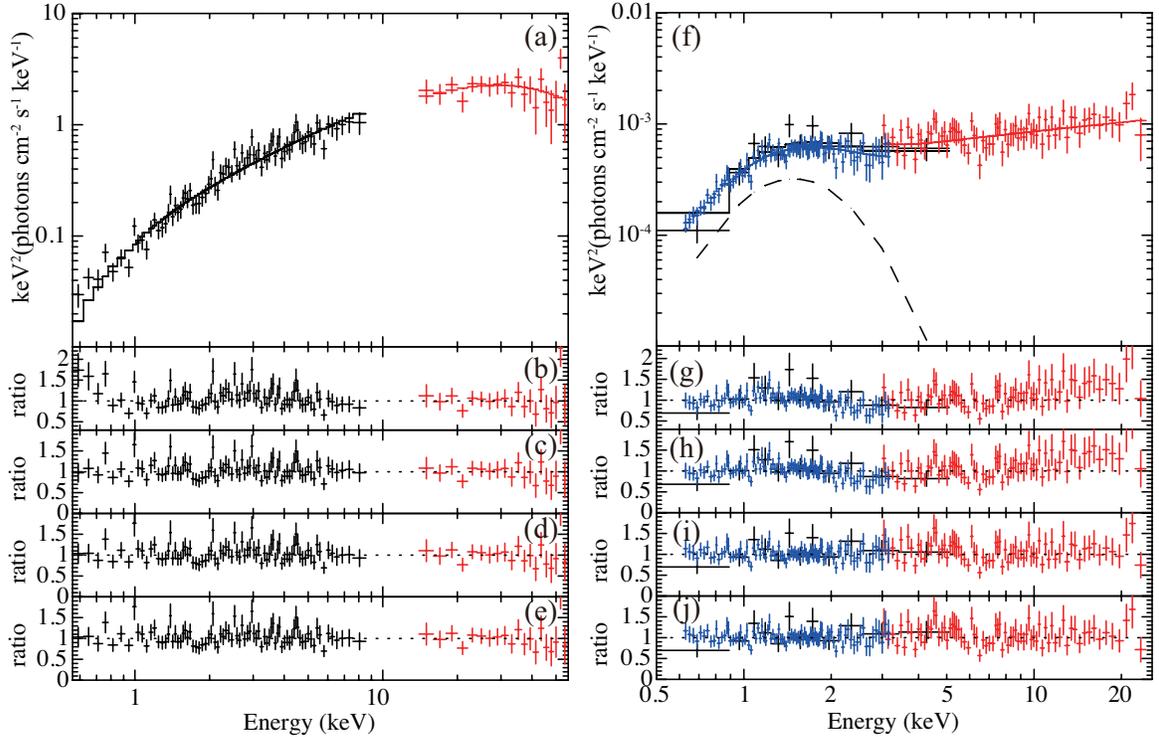}
\caption{Wide-band X-ray spectra obtained at $T \sim 2 \times 10^2$ s
(left) and $\sim 2 \times 10^5$ s (right), fitted with 4 different 
models. The \swift, NICER, and \nustar~data are plotted in black, blue, and red, respectively. The best-fit {\tt TBabs*pcfabs*powerlaw} model and 
{\tt TBabs*(bbodyrad+powerlaw)} model are adopted in the panels 
(a) and (f), respectively. The dashed line in panel (f) presents the contribution of 
{\tt bbodyrad}. The data versus model ratios for {\tt TBabs*powerlaw} (b and g), 
 {\tt pcfabs*TBabs*powerlaw} (c and h), {\tt TBabs*(bbodyrad+powerlaw)} 
 (d and i), and {\tt TBabs*pcfabs*powerlaw} (e and j) are also plotted. 
\label{fig:spec_wideband}}
\end{figure*}

The model successfully reproduced all the spectra.
Figure~\ref{fig:spec_par} plots the individual 
fit parameters in the chronological order. 
The spectrum until $T \sim 10^4$ s had a small 
photon index of $\Gamma =$ 1.0--1.2. The value 
increased in the following rapid decline, 
and became $\Gamma \sim 2.0$ after the decay. 
During the decay, the column density was comparable 
with the total Galactic column, 
except for the plateau phase in $T \sim 3\times 
10^4$--$10^5$ s, where $\nh$ increased up to 
$\sim 1 \times 10^{22}$ cm$^{-2}$. At this phase, 
$\Gamma$ was estimated to be a slightly larger
value, $\Gamma \sim 2$--$3$.
These behaviors may 
suggest variations of the intrinsic absorption, 
although it could be affected by the degeneracy 
with $\Gamma$. At late times of and after the 
decay, the $\nh$ values are consistent with or 
slightly smaller than the Galactic column, 
but have large uncertainties due to the low statistics.

\begin{deluxetable*}{cCCCCCCCCCC}[htb!]
\tablecaption{Best-fit 
parameters for the wide-band spectra. \label{tab:pars}}
\tablecolumns{11}
\tablenum{2}
\tablewidth{0pt}
\tablehead{
\colhead{ID} & \colhead{$\nh$} &
\colhead{$\Gamma$} & \colhead{$E_\mathrm{cut}$} & \colhead{$N$\tablenotemark{a}} & \colhead{$kT_\mathrm{in}$, $kT_\mathrm{bb}$} & \colhead{$R_\mathrm{in}$, $R_\mathrm{bb}$\tablenotemark{b}} &
\colhead{$\nh^\mathrm{pcf}$} & \colhead{$C_\mathrm{v}$\tablenotemark{c}} 
& F_\mathrm{X} &  \colhead{$\chi^2/\mathrm{dof}$}  \\
\colhead{} &
\colhead{$10^{21}$ cm$^{-2}$} &
\colhead{} &
\colhead{keV} &
\colhead{} &
\colhead{keV} &
\colhead{km} &
\colhead{$10^{21}$ cm$^{-2}$} &
\colhead{} &
\colhead{erg s$^{-1}$ cm$^{-2}$} &
\colhead{}
}
\startdata
\multicolumn{10}{l}{{\tt TBabs*cutoffpl}} \\ 
SP1 & 2.4 \pm 0.1 &  0.8 \pm 0.1 & 24^{+7}_{-5} & 0.13 \pm 0.02
& - & -  & - & - & 9 \times 10^{-9} & 119/99  \\ 
SP2 & 2.5 \pm 0.3 &  1.98 \pm 0.09 & > 196 & (7 \pm 1) \times 10^{-4} & -
& - & - & - & 5 \times 10^{-12} & 204/172   \\ \tableline 
\multicolumn{11}{l}{{\tt TBabs*pcfabs*cutoffpl}} \\ 
SP1 & < 1.8 &  1.2^{+0.3}_{-0.1} &  36^{+30}_{-9} & 0.24^{+0.12}_{-0.04} & -
& - & 17^{+13}_{-5} & 0.7^{+0.04}_{-0.15} & 8 \times 10^{-9} & 96/97   \\
SP2 &2.6 \pm 0.2 &  2.0 \pm 0.1 &  > 193 & (7 \pm 1) \times 10^{-4} & -
&- & < 0.25  & -\tablenotemark{e} & 5 \times 10^{-12} & 203/170   \\ \tableline 
\multicolumn{11}{l}{{\tt TBabs*(bbodyrad+cutoffpl)}} \\ 
SP1 & 6.8 \pm 0.2 & 1.0 \pm 0.2 & 30^{+14}_{-8} & 0.19^{+0.04}_{-0.03} 
& 0.07^{+0.02}_{-0.01} & 6^{+12}_{-4} \times 10^3 & -  & - & 8 \times 10^{-9} & 97/97   \\ 
SP2 & 2.1 \pm 0.5 & 1.7^{+0.1}_{-0.2} & > 49  & (4 \pm 1) \times 10^{-4}  &
0.33 \pm 0.05 & 2.0^{+0.8}_{-0.5} & - & - & 6 \times 10^{-12} & 150/170 \\ \tableline 
\multicolumn{11}{l}{{\tt TBabs*(diskbb+cutoffpl)}} \\ 
SP1 & 6.7 \pm 0.2 &  1.0 \pm 0.2 & 31^{14}_{-8} & 0.19^{+0.04}_{-0.03} &
0.08 \pm 0.02 & 8^{+19}_{-6} \times 10^3 & - & - & 8 \times 10^{-9} & 97/97   \\ 
SP2 & 2.6^{+0.5}_{-0.4} & 1.7^{+0.1}_{-0.3} & >33  & (3 \pm 1) \times 10^{-4}  &
0.5 \pm 0.1 & 1.4^{+0.7}_{-0.4}& - & - & 6 \times 10^{-12} & 150/170 \\ \tableline
\enddata
\tablenotetext{a}{Normalization of {\tt cutoffpl}, defined as the photon flux in units of photons keV$^{-1}$ cm$^{-2}$ s$^{-1}$ at 1 keV.}
\tablenotetext{b}{$D=8$ kpc and $i=60^\circ$ (for {\tt diskbb}) were assumed.}
\tablenotetext{c}{Covering fraction of {\tt pcfabs}.}
\tablenotetext{d}{Unabsorbed 1--100 keV flux.}
\tablenotetext{e}{Not constrained.}
\end{deluxetable*}

\subsubsection{Absorption Structure} \label{sec:spec_abs}

In Fig.~\ref{fig:spec}, a possible absorption feature 
was found at 3--4 keV in the \swift~and NICER spectra 
obtained in $T = 2 \times 10^4$--$5 \times 10^4$ s. 
To describe the profile, we tested (1) a Gaussian 
line model ({\tt gauss}, with a negative normalization) 
and (2) a simplified edge model ({\tt edge}). 
The same absorbed power-law model as in Section~\ref{sec:spec_fit} 
was used as the continuum model, where all the parameters 
were allowed to vary. We also adopted the same 
energy ranges of the spectra as those in that section. 

The best-fit values of the {\tt gauss} and {\tt edge} 
components are listed in Table~\ref{tab:pars_abs}, 
and the \swift/XRT and NICER 
data and their best-fit models with {\tt edge} are 
shown in Fig.~\ref{fig:spec_abs}. 
The quality of fit for the \swift/XRT spectrum 
was significantly improved from the case of the 
continuum model, whereas the improvement was not significant 
in the NICER data. Using the script {\tt simftest} on XSPEC, 
we estimated the chance probability of improvement by 
the additional {\tt edge} component as 0.001 and 0.2 
(corresponding $\sim$ 4$\sigma$ and $\sim$ 1$\sigma$) 
for the \swift~and NICER data, respectively. The two 
spectra gave consistent edge/line energies. 
The Fe K line and/or edge around 6--7 keV 
are most prominently seen in many X-ray sources  
including accreting black holes and neutron stars. 
Assuming the observed feature 
originated in the neutral Fe K-shell edge, 
the redshift estimated from the {\tt edge} model 
is as $z \sim 1.1$, (see Section~4.2 for discussion).

If the feature is the neutral Fe K absorption edge of 
cold gas, Fe K emission or absorption lines would also be 
produced, depending on its geometry and temperature. 
We tested to add a negative/positive Gaussian 
component at $z=1.1$ with a line center energy of 
6.4 keV to the absorbed power-law continuum model, 
to consider the neutral Fe K$\alpha$ 
absorption/emission line. The 90\% upper limits of 
the line strengths for the \swift~data 
were estimated to be $4.0 \times 10^{-4}$ 
and $3.3 \times 10^{-4}$ photons cm$^{-2}$ s$^{-1}$ 
($\sim 1.2 \times 10^2$ eV and $\sim$ 97 eV in 
the equivalent width), for the positive and negative 
Gaussian cases, respectively.

We also applied, to the \swift~spectrum, 
the neutral Compton reflection model {\tt pexmon} \citep{nan07}, 
which accounts for both the Fe K emission lines and edge 
self-consistently, to test the 
possibility of a reflection component like those seen 
in X-ray binaries and AGNs. We replaced the power-law component 
in the original continuum model to {\tt pexmon}
and allowed to vary the reflection scale factor ({\tt rel\_refl}),
which is the solid angle $\Omega$ of the reflector visible from 
the illuminating source, in units of $2 \pi$. 
The abundance of Fe and other elements was assumed 
to be the same as the Solar abundance, and the redshift and 
the cutoff energy were fixed at 1.1 and 0 keV 
(i.e., the high energy cutoff was not included). 
We tested the two cases of inclination angles: $i = 0^\circ$ 
and $60^\circ$. In both cases, the model gave 
somewhat better fits than the absorbed power-law model,  
yielding C-stat/d.o.f. $=$ 217/263. We obtained only a weak upper 
limit of the reflection factor, $\Omega/2 \pi < 2.0$ 
and $2.6$, for $i = 0^\circ$ and $60^\circ$, respectively.

\subsubsection{Analysis of Wide-band Spectra} \label{sec:spec_wide_band}

Hard X-ray spectra with good statistics were 
obtained from the \swift/BAT observation at an 
early phase ($T \sim 10^2$ s) and the 
\nustar~observation in the last part of decay 
($T \sim 2 \times 10^5$ s). 
Combining them with the \swift/XRT and NICER data 
taken (quasi-)simultaneously, we obtained wide-band 
spectra in $T \sim 2 \times 10^2$ s (Hereafter SP1) 
and in $T \sim 2 \times 10^5$ s 
(SP2), as shown in Figure~\ref{fig:spec_wideband}(a) 
and (f), respectively. The SP1 spectrum 
was produced from the \swift/XRT and BAT data of 
OBSID$=$00954304000, and the SP2 spectrum was from 
the \swift/XRT, NICER, and \nustar~data of 
OBSID$=$00954304004, 2201010105, and 90601304002, 
respectively. The former spectrum has 
a clear high-energy cutoff at $\sim 30$ keV, while 
the latter shows a flat profile extending to at least 
$\sim 20$ keV. Here, we adopted $\chi^2$ statistics 
for the \swift/BAT spectrum, because it was already 
grouped by channels to have Gaussian distribution 
and hence C-statistics should not be used\footnote{\url{https://swift.gsfc.nasa.gov/analysis/threads/batspectrumthread.html}}. 
For consistency, the $\chi^2$ statistics 
are applied to the \swift/XRT, NICER, and \nustar~spectra 
as well. To ensure that the distrubution underlying the data 
can be approximated by the Gaussian distribution, 
the \swift/XRT data of SP1 
were binned so that at least 40 counts are included in 
each spectral bin. The \swift/XRT, 
NICER, and \nustar~spectra in SP2 were binned to have 
minimum 20, 20, and 40 counts per bin, respectively.

We first fit the two spectra with an absorbed 
cutoff power-law model: {\tt TBabs*cutoffpl} in the 
XSPEC terminology. The model versus data ratios 
are shown in Figure~\ref{fig:spec_wideband}(b) and (g) 
and the best-fit parameters are listed in 
Table~\ref{tab:pars}. Although the model fit 
both spectra fairly well, small residual structures 
remained. In Fig.~\ref{fig:spec_wideband}(b) 
discrepancy between the data and model 
can be seen at lowest energies, 
and in Fig.~\ref{fig:spec_wideband}(g), a small hump 
in 0.8--2 keV and an excess  
above $\sim 10$ keV are present.

Next we investigated if an additional, optically-thick thermal 
component or partial covering absorption can 
better reproduce the soft X-ray profiles. 
We adopted {\tt pcfabs} as the partial covering 
absorption model at $z = 0$. For the thermal component, 
we tested a blackbody component 
({\tt bbodyrad}) and a disk blackbody component 
({\tt diskbb}; \citealt{mit84}). 
In SP1, these models significantly improved the quality 
of fit (Fig.~\ref{fig:spec_wideband}c, d, and e), and 
in any of the models the $\chi^2$ values decreased 
by $\sim 20$ with the degrees of freedom reduced 
by 2 (Tab.~\ref{tab:pars}) from the simple absorbed 
cutoff power-law model. The fit quality of SP2 was
not improved by adding the {\tt pcfabs} 
model (Fig.~\ref{fig:spec_wideband}h), but an additional 
thermal component (Fig.~\ref{fig:spec_wideband}i and j) 
reduced the $\chi^2$ value by $\sim 50$.
We obtained a low temperature and a large radius of 
(disk) blackbody in SP1, while a slightly higher 
temperature and a much smaller radius in SP2.

\subsection{Profile of Flux Decay} \label{sec:lc_fit}

To investigate the profile of the long-term flux decay, 
we analyzed the flux light curve in Fig.~\ref{fig:spec_par}(a), 
obtained in the analysis of the soft X-ray spectra 
based on the absorbed power-law model (see Section~\ref{sec:spec_fit}). 
First, to roughly characterise the decay profile, 
we fitted the data points with a single power-law function 
in terms of the time $T$ from the BAT detection: 
\begin{equation}
F_\mathrm{X} \propto(T-T_\mathrm{peak})^{-\alpha}, \label{eq:lc_fit}
\end{equation}
where $F_\mathrm{X}$ is the unabsorbed 0.5--10 keV flux 
and $T_\mathrm{peak}$ is the time at the flux peak, which we 
assumed to be the second scan of MAXI at $T \approx 
2 \times 10^3$ s. Here, 
we only considered the decay period from $T \approx 
1 \times 10^3$ s to $2 \times 10^4$ s and ignored 
the initial flux rise, the final constant-flux phase, 
and the plateau phase in $T \approx$ (2--5) $\times 
10^4$ s during the decay. A decay index of 
$\alpha \approx 1.7$ was found to approximately 
reproduce the observed decay.

Next, we applied a multiply broken power-law model
to the flux data over the entire period, to better 
describe the decay profile and estimate the total 
radiated energy accurately. We found that the flux 
profile can be well reproduced by 
using 8 power-law segments with a photon index of $\alpha_1$, $\alpha_2$, ..., 
and $\alpha_8$, and with 7 breaks: 
$T_\mathrm{br1}$, $T_\mathrm{br2}$,
..., and $T_\mathrm{br7}$, from earlier to later times.
We described the initial flux rise until the 
second MAXI scan by two power-law segments. 
Because only 3 data points were available in this 
period, we fixed $\alpha_1$, and $\alpha_2$ at the values 
calculated from them and assumed $T_\mathrm{br1}$
and $T_\mathrm{br2}$ as the times of first and 
second MAXI scan. 
We note that this complex model is not physically 
motivated and is used just to obtain an accurate 
measurement of the fluence.

\begin{figure}[ht!]
\plotone{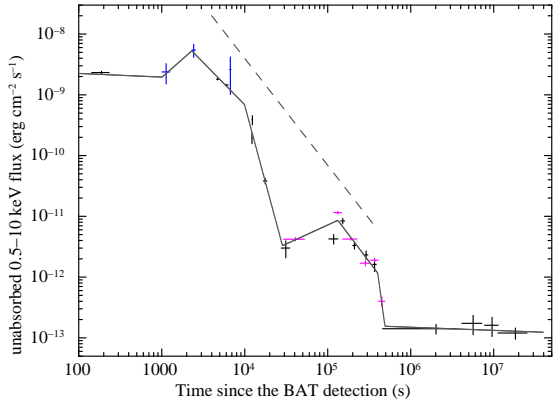}
\caption{Flux light curve (same as Fig.~\ref{fig:spec_par}a) 
and the best-fit multiply broken 
power-law model. The same data colors as Fig.~\ref{fig:spec_par}(a) 
are adopted. The dashed line indicates a single power-law with 
an index of $\alpha = 1.7$ in terms of the time from the flux 
peak (see Equation~\ref{eq:lc_fit}). 
}
\label{fig:lc_fit}
\end{figure}

Figure~\ref{fig:lc_fit} plots the light curve data 
and the best-fit model 
and Table~\ref{tab:lc_fit} 
gives the best-fit parameters. 
We also estimated the X-ray fluence in 0.5--10 keV for the 
total outburst period as $F_\mathrm{0.5-10~keV} = 2.9 \times 
10^{-5}$ erg cm$^2$, by integrating the best-fit multiply 
broken power-law model in Fig.~\ref{fig:lc_fit} 
over the entire period from $T=0$ s to $3 \times 10^7$ s.

In Section~\ref{sec:spec_wide_band}, we found that 
including a partial covering absorption component 
give a better fit to the broadband X-ray spectrum 
at the early phase. To investigate if its inclusion 
affect the flux light curve, we tested to add {\tt 
pcfabs} to the {\tt TBabs*powerlaw} model and fit the 
soft X-ray spectra in Section~\ref{sec:spec_fit}. 
The {\tt TBabs*pcfabs*powerlaw} model improved 
the fits of the spectra obtained 
by the end of the first rapid decay ($T < 10^5$ s) 
but not in the later period. The unabsorbed 0.5--10 keV 
fluxes were found to change only by $\lesssim 10$\% from 
the {\tt TBabs*powerlaw} model, and this does not 
affect the overall light curve profile in Fig.~\ref{fig:lc_fit}.

\begin{deluxetable}{CcC}[ht!]
\tablecaption{Best-fit parameters of the multiply broken power-law model for the long-term light curve. \label{tab:lc_fit}}
\tablecolumns{3}
\tablenum{3}
\tablewidth{0pt}
\tablehead{
\colhead{Parameter} & \colhead{Unit} &
\colhead{Value} }
\startdata
\alpha_1 &  & 0.06~\mathrm{(fixed)} \\ 
T_\mathrm{br1} & s & 1.1 \times 10^2~\mathrm{(fixed)} \\ 
\alpha_2 &  & -1.2~\mathrm{(fixed)} \\ 
T_\mathrm{br2} & s & 2.5 \times 10^3~\mathrm{(fixed)} \\ 
\alpha_3 &  & 1.4 \pm 0.1 \\ 
T_\mathrm{br3} & s & (1.0 \pm 0.1) \times 10^4 \\ 
\alpha_4 &  & 5.1^{+0.8}_{-0.7} \\ 
T_\mathrm{br4} & s & 2.9^{+0.3}_{-0.2} \times 10^4 \\ 
\alpha_5 &  & -0.6 \pm 0.1 \\ 
T_\mathrm{br5} & s & (1.30 \pm 0.05) \times 10^5 \\ 
\alpha_6 &  & 1.8 \pm 0.2 \\ 
T_\mathrm{br6} & s & > 3.8 \times 10^5 \\ 
\alpha_7 &  & >6 \\ 
T_\mathrm{br7} & s & 4.8^{+0.5}_{-0.3} \times 10^5 \\ 
\alpha_8 &  & 0.1^{+0.2}_{-0.1} \\ 
\tableline 
F_\mathrm{0.5-10 keV}\tablenotemark{a} & erg cm$^{-2}$ & 2.9 \times 10^{-5} \\ \tableline 
\enddata
\tablenotetext{a}{0.5--10 keV fluence in the period of $T=0$--$3 \times 10^7$ s.}
\end{deluxetable}

\section{Discussion}
\subsection{Summary of X-ray Properties}

We have studied the X-ray properties of \srcname~using 
\swift, MAXI, NICER, and \nustar~data.   
Here we summarise the results, to discuss the 
nature of the source in the following sections.

\begin{itemize}
\item A rapid outburst decay by $\sim 5$ orders of magnitude 
within $\sim 5$ days was observed. After the decay, the source 
flux was almost constant over 9 months (Fig.~\ref{fig:longterm_lc_hid}). 

\item The overall decay profile can be approximated with the function 
$\propto t^{-1.7}$, where t is the time from the flux peak. More precisely, 
the profile is better described with a multiply broken power-law model 
(Fig.~\ref{fig:lc_fit} and Tab.~\ref{tab:lc_fit}). The total X-ray 
fluence until $3 \times 10^7$ s was estimated to be $\sim 3 \times 10^{-5}$ 
erg cm$^{-2}$.  

\item We detected strong short-term variations on time scales 
of 1--$10^3$ s were seen during the decay (Fig.~\ref{fig:lc_short}). 
A possible periodicity of 8.96 s was detected in the first \swift/XRT 
observation, but not in the other observations (Fig.~\ref{fig:powspec_sw} and Fig.~\ref{fig:folded_lc}).

\item The soft X-ray spectrum was well described with an 
absorbed power-law model. The source showed the hardest 
spectrum with a photon index of $\sim 1$, which gradually 
increased with the decreasing flux and finally reached 
$\Gamma \sim 2$ after the decay (Fig.~\ref{fig:spec_par}). 
The absorption column density was comparable to the total Galactic column, 
although some variations were present.

\item A clear spectral rollover was observed at $\sim$ 30 keV 
at the initial outburst rise. The wide-band X-ray spectrum 
at this phase was better described by adding a partial 
covering absorption or a (disk) blackbody component
to an absorbed cutoff power-law model. An additional 
thermal component also improved the fit of the wide-band 
spectrum taken in the last period of the decay (Fig.~\ref{fig:spec_wideband} 
and Tab.~\ref{tab:pars}).

\item A possible absorption feature at $\sim 3.4$ keV 
was detected in $T \sim 2 \times 10^4$--$5 \times 10^4$, 
when the light curve exhibited a plateau following 
the first steep decay (Fig.~\ref{fig:spec}, \ref{fig:spec_abs},
and Tab.~\ref{tab:pars_abs}). 
Assuming it as the neutral 
iron-K edge, the redshift is estimated to be $z = 1.1$.

\end{itemize}
\subsection{Possible Nature of the Source}

\subsubsection{A Galactic Origin}
Let us first investigate the possibility of a Galactic transient. 
Given that the source is located near the Galactic plane, 
it would be reasonable to take this possibility into account. 
Because the absorption column density is comparable 
to the total Galactic column for a large fraction of 
the outburst period, the source would not have a very 
close distance. Assuming an isotropic emission, 
the maximum and minimum luminosities in 0.5--10 keV 
are estimated as $2 \times 10^{37} (D/8~\mathrm{kpc})^2$ 
erg s$^{-1}$ and $7 \times 10^{32} (D/8~\mathrm{kpc})^2$ 
erg s$^{-1}$, respectively. The large luminosity range 
and the short time scale of the decay, and the  
relatively faint optical counterpart (with an apparent 
$r'$ band magnitude of $\sim 20$ mag; \citealt{maz20}) 
rule out the possibility of a high mass X-ray binary 
\citep[e.g.,][]{wal15}. The timescale of the 
X-ray decay is much shorter than magnetar outbursts 
\citep[$\sim 100$ days;][]{cot18}, although a couple of 
them showed two-step decays reminiscent of that seen in 
our target.
The observed hard, non-thermal spectrum without any prominent 
emission lines also makes unlikely the possibilities of 
a stellar flare \citep{gud09} and a CV \citep{muk17}. 

We have confirmed the 8.96-s peak reported by \citet{ken20} 
in the periodogram made with the first \swift/XRT observation.
Given the fact that it did not appear in any other observations, 
the variation was either a transient phenomenon of the source 
or an artefact (such as red noise). If the former is the case, 
it is difficult, due to the limited exposure, to determine 
whether it is a pulsation or a quasi-periodic oscillation (QPO). 
Hence, the possibilities of an X-ray binary pulsar and 
a magnetar are not proved nor ruled out.

Typical black hole (BH) and neutron star (NS) low mass 
X-ray binaries (LMXBs) are also difficult to explain 
some of the observed properties of \srcname. 
The luminosity range of the decay is 
consistent with LMXBs, and the upper limit of the radio flux, 
18 $\mu$Jy at 7.5 Hz, obtained 5 days 
after the discovery \citep{bor20}, also agrees 
with the radio/X-ray flux correlations of both BH and NS 
LMXBs \citep[e.g.,][]{cor13}. 
However, the observed decay period of $\sim$ 5 days 
is much shorter than those of typical LMXBs: 
a few tens of days to more than a year. This means that 
\srcname~accreted much smaller mass than normal LMXBs 
in an unusually short period. Moreover, the spectrum 
in the outburst rise had a somewhat small photon index, 
$\Gamma \sim 1.0$, compared with typical LMXBs in 
the hard state, $\Gamma \sim 1.5$--1.9 \citep[e.g.,][]{don07}. 

Yet these unusual properties cannot completely rule out  
a LMXB nature. There are actually some LMXBs that show
similar properties. An LMXB candidate showing such a 
short outburst with a similar luminosity range 
has recently been found \citep[MAXI J1957$+$032;][]{ber19}.
There are also several known LMXBs with short outbursts 
but smaller peak luminosities of $10^{34}$--$10^{36}$ 
erg s$^{-1}$, which are called ``very faint X-ray transient'' 
\citep[VFXT;][]{wij06,deg10,bah20}. 
In addition, very hard spectra in bright hard 
state have been observed in a couple of NS LMXBs \citep{par17}
and BH LMXBs including V 404 Cyg \citep[e.g.,][]{kim16}, 
V4641 Sgr \citep[e.g.,][]{mai06}, 
and Swift J1858.6$-$0814 \citep{har20}.
All these sources have a strong short-term flux 
variation by a more than order of magnitude 
\citep[e.g.,][]{kit89, rev02, wij17}, 
also in agreement with the present case. 
In addition, partial covering absorption is 
sometimes required in these sources 
especially at high flux phases \citep{kim16}, 
consistent with \srcname~at a bright phase.

We note that the photon index of \srcname~after 
the decay, $\lesssim 2$ is reminiscent of a BH 
rather than a NS LMXB, which usually shows a 
larger value, $\Gamma \sim 3$ \citep{wij15}, 
although the NS LMXB EXO 1745$-$248 showed a similar 
hard spectrum in quiescence \citep{riv18}. 
On the other hand, a NS LMXB would rather be favored, 
in that a very small radius of the thermal component, 
1--2 (D/8 kpc) km, consistent with emission from 
a NS surface, was obtained from the \nustar$+$\swift/XRT$+$NICER 
spectrum at a faint phase. This holds even if 
the observed power-law component was entirely 
produced by Comptonized disk photons.
Assuming conservation of the number of photons and 
an isotropic scattered emission \citep{kub04},  
and using the intrinsic disk photons estimated 
from the best-fit {\tt diskbb+cutoffpl} model, 
we obtain the inner disk radius as $\sim 3$ (D/8 kpc) km 
(where $i=60^\circ$ is assumed). 
Similar thermal emission likely emitted by the 
NS surface can be seen in NS LMXBs at low X-ray 
luminosities below $\sim 10^{35}$ erg s$^{-1}$ 
 but not at a more luminous hard state, where the power-law 
component dominates the X-ray flux \citep[e.g.,][]{sak14,shi17}. 
The thermal 
component in \srcname~was observed at a consistent 
luminosity; the second rapid decay from $10^5$ s 
since the discovery, in which this component 
was found, started at $\sim 10^{35}$ ($D/8$ kpc)$^2$ 
erg s$^{-1}$.

Even if an LMXB scenario works for all the properties 
raised above, the observed absorption feature is 
difficult to explain. Although ionized Si and S lines 
from disk winds are sometimes seen at similar energies 
\citep[e.g.,][]{ued09}, they are much 
weaker than the present case and always with 
strong iron K $\alpha$ lines at $\sim 7$ keV.
Assuming the observed feature as the redshifted Fe 
K edge, we obtained $z = 1.1$. If this is the 
gravitational redshift caused by a Schwarzschild 
stellar mass BH, 
the feature should be produced at $\sim$ 1.3 
$R_\mathrm{S}$, where $R_\mathrm{S}$ is the 
Schwarzschild radius. It is unclear that an absorber 
with such a small size can exist in the vicinity of 
a BH. Another possibility is that the observed feature 
is a different blueshifted line produced by a relativistic outflow 
in an relativistic speed, like baryonic 
jets observed in SS 433. However, it is unclear 
if such baryonic jets can exist at a low Eddington 
ratio and produce a single absorption feature.

Instead, the origin of the source might be 
explained by a tidal disruption of an asteroid 
by a neutron star \citep{new80,col81}, 
which was considered for the origin of 
GRBs 30--40 years ago.
In this model, the collision of an asteroid 
and a NS surface produces strong MeV 
gamma-rays and hot plasma. The plasma 
could be rapidly cooled by radiation 
and cause Compton down scattering, 
resulting in a power-law shaped X-ray spectrum.
The radiation energy of \srcname~released in $10^7$ s 
is estimated as $2 \times 10^{39} (D/8~\mathrm{kpc})^2$ 
erg, which is converted to a total accreted mass of  
$\sim 10^{20}$ $(\eta/0.01)$ $(D/8~\mathrm{kpc})^2$ g (where $\eta$ is 
the radiative efficiency), consistent with an asteroid.
It is unclear, however, whether or not this scenario can 
explain the observed evolution of photon index of the X-ray 
spectrum and the absorption feature.

\subsubsection{An Extragalactic Origin}

We next investigate the possibility of an extragalactic origin. 
If the source is an extragalactic X-ray binary located in 
a nearby galaxy, the maximum luminosity is far beyond the 
Eddington luminosity of a BH with a mass of 10 $M_\sun$. 
Even if the distance is as small as $\sim 1$ Mpc, the peak  
luminosity is calculated as $\gtrsim 10^{41}$ erg s$^{-1}$. Such a bright 
X-ray binary is categorized as a hyper luminous X-ray source 
and considered to have an intermediate mass BH \citep{far09}. 
In this case, it is even more difficult, than in the case of 
a normal X-ray binary with a stellar mass BH, to explain 
the short duration of the outburst, because of the large 
accretion disk size. For the same reason, an AGN origin 
is unlikely. 

Some of the observed properties of \srcname~agree with those 
of GRBs. Absorption features have been detected in a few GRBs 
\citep{ama00, fro04, bel14} and interpreted as the redshifted 
iron K line or edge, although they were detected mainly in 
the prompt phase, unlike the present case (but see \citealt{bel14}). 
The absorption feature in our case was observed 
just after the first rapid decay, which may be considered 
to be the beginning of the afterglow, 
and could be explained by the interaction between the jets 
and the surrounding gas.
If the estimated redshift $z = 1.1$ is the cosmological 
redshift, it is converted to a luminosity distance of 7 Gpc, 
assuming $H_0 = $ 70 km s$^{-1}$ Mpc$^{-1}$, 
$\Omega_\mathrm{M} = 0.3$, and $\Omega_\mathrm{\Lambda} = 0.7$. 
In this case, the maximum and minimum luminosities 
are calculated to be $1 \times 10^{49}$ erg s$^{-1}$ and 
$4 \times 10^{44}$ erg s$^{-1}$, 
respectively, where isotropic emission is assumed. 
The redshift and maximum luminosity are well within 
the distribution of known GRBs. The source showed 
a hard spectrum with $\Gamma \sim 1$ and a spectral 
rollover at $\sim 30$ keV in the early phase, 
which are compatible with soft GRBs in the prompt phase
\citep[e.g.,][]{yon10}. 
Using the best-fit cutoff power-law model of 
the \swift/XRT+BAT spectrum (see 
Section~\ref{sec:spec_wide_band}), we obtain the 
2--10 keV luminosity of $5 \times 10^{48}$ erg 
s$^{-1}$ in the rest frame of the source. 
This is consistent with the correlation 
between the peak energy and X-ray luminosity 
of \swift-detected GRBs reported by \citet{dav12}.

The decay profile for the first $10^6$ s  
and the spectral evolution of \srcname~also share 
similar properties with GRBs. The two-step decay 
could be attributed to the prompt and afterglow 
emission. 
However, the oberved plateau phase 
started later and lasted longer by one 
order of magnitude than the 
typical shallow decay phase of GRBs \citep{zha06}.
Moreover, the constant flux over 9 months after 
the decay is very different from, and difficult 
to explain by, X-ray afterglows of GRBs.

Instead, the source may be interpreted as 
a TDE by a supermassive BH. 
The overall decay profile is approximated 
to $t^{-1.7}$, which is consistent with 
what is theoretically 
expected for a TDE \citep[$t^{-5/3}$;][]{ree88,phi89} 
and the actual X-ray light curves 
of observed TDEs \citep[e.g.,][]{bur11, che12}. 
The strong short-term variability is reminiscent 
of the jetted TDE Swift J164449.3$+$573451 \citep[][hereafter Swift J1644]{bur11, man16}.
X-ray Absorption features have also been detected in a TDE 
at lower energies than that in the present case \citep{mil15}, 
and were interpreted as an outflow at a velocity of a few 
hundred km s$^{-1}$. The total radiation energy in 1--20 keV  
is calculated from the fluence to be $9 \times 10^{52}$ erg 
in the rest frame of \srcname, assuming an isotropic emission
and a luminosity distance of 7 Gpc. 
This corresponds to a total 
accreted mass of $\sim 0.9$ $M_\sun$, assuming the radiative 
efficiency of the standard accretion disk of a Schwarzschild 
BH. The peak luminosity, $\sim 10^{49}$ erg s$^{-1}$ is more 
than one order of magnitude larger than that of the most 
luminous TDE Swift J1644. 

Swift J1644 exhibited a steep flux drop $\sim$ 500 
days after the discovery and an almost constant flux for 
the following $\sim 1000$ days, which can be explained by 
the end of accretion and scattering of X-rays with the 
surrounding gas, respectively \citep{che16}. 
Similar flux behaviors were observed in \srcname, 
with a rapid decay 
until $T \sim 5 \times 10^5$ s and a following 
constant-flux phase. If the end of the decay at $\sim 5 
\times 10^5$ s actually reflects the end of accretion, 
the BH mass ($\mbh$) of \srcname~can be estimated 
by comparing the decay time, or the accretion 
period, of the two sources, as described below. 
We note that the scattering echo scenario for 
the constant-flux phase requires spectral 
softening, because of the energy dependence 
of the scattering angle. However, no significant 
softening was observed and therefore a different 
explanation would be needed in the present case. 

When a star is tidally disrupted by a BH, 
some of the disrupted matter orbits around it 
and falls onto the BH \citep{ree88}. 
The kinetic energy of the disrupted matter 
in the unit mass is described as 
\begin{equation}
    \epsilon \sim \frac{G\mbh}{a},
\end{equation}
where $a$ and $G$ are the semi major axis and the 
Gravitational constant, respectively. 
Combining the Kepler's third law, 
$T \sim \sqrt{a^3/G\mbh}$, we obtain 
\begin{equation}
    T \propto \mbh \epsilon^{-\frac{3}{2}}.
\end{equation}
Thus, the accretion time is proportional to $\mbh$. 

Using this relation, the BH mass of \srcname~should 
be $\sim 10^2$ times smaller than that of Swift J1644, 
which was previously estimated as $10^6$--$10^7$ 
$M_\sun$ \citep{bur11}. 
In this case, we obtain a BH mass of \srcname~of 
$10^4$--$10^5$ $M_\sun$, and therefore
we may have witnessed a rare case of a TDE in 
a dwarf galaxy. The smaller BH mass and higher peak 
luminosity than those of Swift J1644 imply that 
\srcname~is a more highly beamed, jetted TDE. 
This is of course just a rough estimate, and 
different approaches such as observations of 
the host galaxy would be required for a more 
accurate measurement.

A QPO at 5 mHz was detected in Swift J1644 
and associated with the Keplerian frequency at 
the innermost stable circular orbit of the BH 
\citep{rei12}. If the observed 8.96 s variation 
in \srcname~is explained by a QPO produced 
in the same manner, its BH mass is calculated 
to be $10^4-10^5 M_\sun$, depending on the BH 
spin. This is consistent with the mass 
estimated above from the decay period.

\section{Summary and Future Prospects}

Using X-ray data from MAXI, \swift, NICER, 
and \nustar, we have found unusual X-ray 
properties of \srcname~that are 
difficult to explain by any of known 
class of objects. A plausible
interpretation may be an LMXB with multiple 
extreme properties, or a 
tidal disruption of a star 
by a supermassive BH or an asteroid by 
a Galactic NS. 
In any case, this source has provided an 
opportunity to study an interesting and perhaps 
rare astrophysical phenomena around a compact object. 

To uncover the true nature of the source, 
observations of the optical/infrared 
counterpart are essential. So far, 
only several results from photometric observations 
have been reported, all of which were performed 
in the first couple of days after the discovery. 
Future high sensitivity and spatial 
resolution optical imaging, using largest class 
telescopes, could provide morphology information 
of the optical counterpart and determine its 
host galaxy if it is an extragalactic object.
In addition, high sensitivity spectroscopy  
would enable an accurate determination of 
the redshift of the source.

\acknowledgments
This work made use of \swift~data supplied by the 
UK Swift Science Data Centre at the University of Leicester, 
and MAXI data provided by RIKEN, JAXA and the MAXI team. 
We thank the \nustar~team for performing the ToO observation.
MS thanks Yuichi Terashima, Tohru Nagao, and Kazuma Joh, 
for discussion of optical and infrared counterparts.
Part of this work was financially supported 
by Grants-in-Aid for Scientific Research 19K14762 (MS) 
from the Ministry of Education, Culture, Sports, 
Science and Technology (MEXT) of Japan. 
PAE acknowledges UKSA support.



\facilities{MAXI~(GSC), \swift~(BAT and XRT), NICER~(XTI), \nustar~(FPMA and FPMB)}


\software{
XSPEC (v12.9.0n; \citealt{arn96}), HEAsoft (v6.26.1; \citealt{heasoft}))
          }



\vspace{6cm}
\bibliography{swj0840p7_ms}{}
\bibliographystyle{aasjournal}



\end{document}